\documentclass[runningheads]{llncs}
\usepackage{graphicx}%
\usepackage{multirow}%
\usepackage{amsmath,amssymb,amsfonts}%
\usepackage{mathrsfs}%
\usepackage[title]{appendix}%
\usepackage{xcolor}%
\usepackage{textcomp}%
\usepackage{manyfoot}%
\usepackage{booktabs}%
\usepackage{algorithm}%
\usepackage{algorithmicx}%
\usepackage{algpseudocode}%
\usepackage{listings}%
\usepackage[utf8]{inputenc}
\usepackage[T1]{fontenc}
\usepackage{url}
\usepackage{graphicx}
\usepackage{subfigure}
\usepackage{psfrag}
\usepackage{listings}
\usepackage{amsmath}
\usepackage{amsfonts}
\usepackage{supertabular}
\usepackage{array}
\usepackage{tabularx}
\usepackage{boldline} %paczka z grubymi liniami
\begin{document}
\title{Transfer of knowledge among instruments in automatic music transcription}

%\author{Anonymous}

\author{Michał Leś\inst{1}\orcidID{0000-0001-7688-1025} \and
Michał~Woźniak \inst{1}\orcidID{0000-0003-0146-4205}}
\authorrunning{M. Leś and M. Woźniak}

\institute{Wrocław University of Science and Technology, Wrocław, Poland \email{michal.les@pwr.edu.pl}}
%\author*[1]{\fnm{Michał} \sur{Leś}}\email{michal.les@pwr.edu.pl}

%\author[1]{\fnm{Michał} \sur{Woźniak}}\email{michal.wozniak@pwr.edu.pl}\orcidID{0000-1111-2222-3333}

%\affil*[1]{\orgdiv{Department of Systems and Computer Networks}, \orgname{Wroclaw~University~of~Science~and~Technology}, \orgaddress{\city{Wrocław}, \country{Poland}}}

\maketitle

\begin{abstract}
    Automatic music transcription (AMT) is one of the most challenging tasks in the music information retrieval domain. It is the process of converting an audio recording of music into a symbolic representation containing information about the notes, chords, and rhythm.
Current research in this domain focuses on developing new models based on transformer architecture or using methods to perform semi-supervised training, which gives outstanding results, but the computational cost of training such models is enormous.

This work shows how to employ easily generated synthesized audio data produced by software synthesizers to train a universal model. It is a good base for further transfer learning to quickly adapt transcription model for other instruments. Achieved results prove that using synthesized data for training may be a good base for pretraining general-purpose models, where the task of transcription is not focused on one instrument.
\end{abstract}

\section{Introduction}
Automatic music transcription (AMT) is a challenging task in the musical information retrieval domain. It is the process of converting music recordings into symbolic representations such as music sheets or MIDI files. The problem is especially visible for polyphonic music, where multiple frequencies affect each other, giving a result which is hard to estimate using simple algorithms for time-frequency analysis.  
%TODO napisać motywację - DONE
%TODO rozwinąć i zakończyć listą osiągnięć - DONE
Current works often use datasets, which are hard to obtain and maintain. For instruments without any electronic interface, the possibility of gathering an extensive dataset containing real-world recordings with transcription is limited. 

This work focuses on discovering the potential of using a model trained on synthesized data for automatic music transcription. These data are easy to generate and could be produced on demand. The intuition behind this idea is related to the possibility of universally recognizing pitch and rhythm by the human ear - it performs similarly no matter the kind of audio source. Synthesized instruments may contain fewer noises than the audio recorded via microphone. However, a model trained on such data may focus more on frequency analysis tasks, which is a good base for fine-tuning it to real-world data.  

%In this work w
During the presented research, we trained the U-net model with BiLinear LSTM on software-synthesized samples containing timbre from different instruments. We improved the model's performance in the target domain based on real-world recordings that prove that the model trained on synthesized instruments could generalize transcription in note and frame metrics across different datasets after zero-shot transfer. We may use it for quick adaptation to another real-world recording dataset competing with models pre-trained on recordings of the specific instrument.

\section{Related Works}
%TODO wprowadzenie - DONE
This chapter focuses on the description of currently used models for transcription and audio processing methods. Later we describe data representation and metrics used in the existing literature.  
%\cite{hawthorne2021sequence}
Current research often focuses on developing other models \cite{benetos2018automatic}, and methods \cite{cheuk2021effect} to improve standard automatic music transcription metrics or find a way to perform unsupervised \cite{maman2022unaligned} or semi-supervised \cite{cheuk2021reconvat} automatic transcription. It may result in analyzing different timbres and instruments less underlined. The importance of timbre in automatic music transcription was presented by Hernandez-Olivan et al. in \cite{hernandez2021comparison}, where they proved that for different instruments, the estimation of source frequency f0 might be a challenging task. In \cite{cheuk2020impact}, authors demonstrated that neural network-based music transcription depends on audio input representation because different spectrograms may pass different information. Modern deep learning models offer the possibility to perform transcription for multiple instruments together with track separation \cite{gardner2021mt3}. Multi-instrument AMT was considered in \cite{wu2020multi}, and results proved that model capable of performing transcription on one instrument is not working well for another instrument, especially when this knowledge is not passed a priori.

%TODO - prebudować tekst, tak żeby pasował do otoczenia - DONE
After the significant development of machine learning models capable of resolving many problems and approximating complex solutions, it was clear that it may be used for automatic music transcription. Most currently existing solutions are based on image representation of audio recordings in a time-frequency domain called spectrograms. Two kinds of spectrograms are used as input for existing deep learning models to visualize traits of frequencies present in audio recordings - Mel spectrogram and CQT.

%\cite{brown1992efficient}
Mel spectrogram is based on the result of time-frequency analysis of sound using Short Time Fourier Transform (STFT) \cite{allen1977short} and proper alignment using a particular "mel scale", which makes the distance between two pitches equal as the listener perceives it. It is mainly used to maximize transcription accuracy, as frequencies are the essential input to train the model for AMT. Constant Q Transform (CQT) \cite{brown1991calculation} is another method to achieve images containing information about played frequencies. Its properties fit well to analyze the music and expose musical information like timbre, which may be crucial for this research. We decided to use CQT as the primary spectrogram function in our experiments.

MIDI files often represent labels in datasets, which contain all information needed for a musician or software musical instrument to reproduce a similar sound. Usually, it may be represented as a table, where each row contains four parameters:
\begin{itemize}
    \item value of pitch - it is a general value used to distinguish between different musical tones,
    \item value of velocity - this value refers to the speed or intensity with which a note or sound is played; it may be often referred to as a measure of how hard a musician strikes a key on a keyboard or plucks a string on a guitar,
    \item note start - it is the time when the tone is started to be played by an instrument,
    \item note end - it is time the tone is ended to be played by an instrument.
\end{itemize}

Timbre is another essential trait of music recordings. The same pitch played by different instruments may result in a different reception of the same tone. It is determined by combining many factors, including a sound's harmonic content, sustain, attack and decay, and how different frequencies are emphasized or de-emphasized. It results in problems with transferring this knowledge because models trained on spectrograms for guitar recordings will not learn the same way of extracting pitch as the model trained on piano recordings.

Another problem with analyzing different timbres is related to available datasets. Only a few datasets contain real recordings, especially for instruments not available in electronic forms, like a piano. Audio data synthesis is present in almost every currently used dataset for AMT \cite{hawthorne2018enabling}, \cite{emiya2010maps}, \cite{xi2018guitarset} - not only for the generation of new recordings but also for validation of real-world samples. The enormous dataset containing partially generated data is MAESTRO \cite{hawthorne2018enabling}, where authors proposed the "Wave2Midi2Wave" approach and created a set of valuable recordings containing traits of real-world ones. There is also some research where authors try to use a semi-supervised approach for transcription to mitigate the problem of small real-world datasets \cite{cheuk2021reconvat} and create the framework for continual learning of different instruments.

Also, currently used metrics for AMT (defined in \cite{raffel2014mir_eval}) are not always enough to measure all aspects of music transcription. Simonetta et al. in \cite{simonetta2022perceptual} proposed another metric, which was correlated with perceptual measures in the outcome of the automatic music transcription model. Due to the lack of timbre generalization capability in the currently existing model, we aim to find a way to create a model which performs well in a new domain (which is related to different instruments) after zero-shot \cite{radford2021learning} transfer of weights.

\section{Proposed approach}
\begin{figure}
    \centering
    \includegraphics[width=12cm]{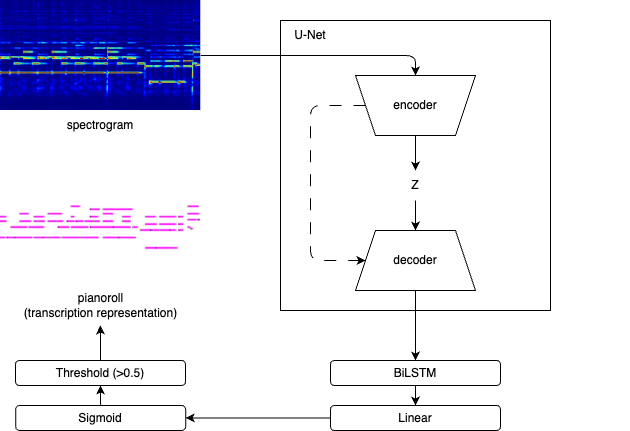}
    \caption{Architecture of base model used for experiments.}
    \label{fig:model}
\end{figure}
As a base for our experiments, we used part of the model presented as a transcriber in \cite{cheuk2021effect} by Cheuck et al. for investigating the spectrogram reconstruction effect. It is composed of U-net \cite{ronneberger2015u}, BiLSTM and a simple linear layer for classification. We decided to use that model for transcription because of a small number of parameters and relatively good results with the possibility to catch time-series dependencies via bidirectional LSTM. Similar architecture was presented by Hawthorne et al. in \cite{hawthorne2017onsets}, but it contained convolutional networks instead of U-net. The exact model configuration is presented in figure \ref{fig:model}.  

%TODO add figure with pipeline - DONE
\begin{figure}[ht]
    \centering
    \includegraphics[width=12cm]{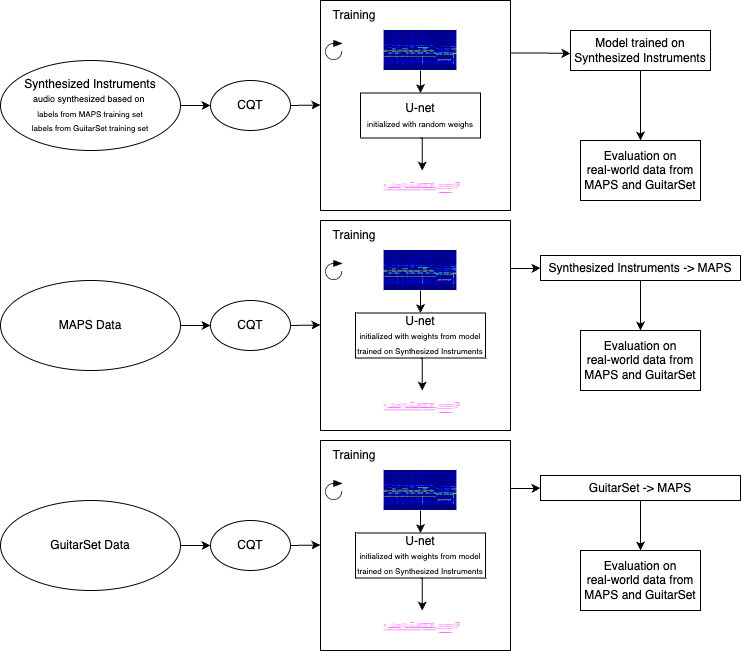}
    \caption{Pipeline for training based on synthesized instruments and evaluation of results.}
    \label{fig:pipeline}
\end{figure}

We proposed a pipeline presented in figure \ref{fig:pipeline} to measure the usefulness of the synthesized data. The data in experiments are randomly sampled during training to extract fixed sequence length from larger tracks, so the original distribution of data should not significantly impact training. For synthesized data generation we used training labels from GuitarSet and MAPS datasets. After audio synthesis, we performed training of the model (resulting in a model trained on Synthesized Instruments - MTOSI) and later used it for fine-tuning to MAPS and GuitarSet data. To compare transfer from MTOSI to transfer from the model trained on each dataset a similar process was performed for the model trained on MAPS dataset and fine-tuned on GuiterSet and vice-versa.

\section{Experiment}
% TODO - add research questions - DONE
The experiments were designed to answer the following research questions:
\begin{itemize}
    \item RQ1: How does the U-net model trained on synthesized data perform on real-world recordings?
    \item RQ2: How does knowledge transfer from the U-net model trained on synthesized data impact training on real-world recordings?
    \item RQ3: Is the U-net model trained on synthesized data a better candidate for fine-tuning on other instruments than the U-net model trained on real-world piano recordings?
\end{itemize}
\subsection{Setup}

% Datasets
\subsubsection{Datasets}
This section describes types of datasets taken into account with descriptions and conditions used in experiments.
\paragraph{MAPS}
MAPS \cite{emiya2010maps} is a commonly used dataset for AMT and contains audio recordings with corresponding music notation in MIDI format or text file containing note onset time, note offset time and value of pitch. It consists of 238 music recordings performed in a different environment and in another way. Part of the data was gathered using Diskclavier piano by recording the automatically playing instrument simultaneously by two omnidirectional Schoeps microphones. Some data was generated using a software-based solution (Steinberg's Cubase SX). It contains music pieces, usual chords, isolated sounds and random sound combinations. For experiments, we used only whole music pieces, which were split into training (80\%), validation (10\%) and testing (10\%) sets. Each recording in MAPS was sampled with a 16 kHz sample rate and transformed into the CQT spectrogram.

\paragraph{GuitarSet}
GuitarSet \cite{xi2018guitarset} is the most popular dataset containing recordings of guitar annotated semi-unsupervised via a hexaphonic pickup. Authors provide annotations in a convenient JAMS format, which may be easily converted to the pitch value with the corresponding onset and offset time. Additionally, pieces of music were played by different musicians and gathered by different microphones, resulting in seven audio channels. For the experiment, recordings with annotations were split into training (80\%), validation (10\%), and testing (10\%) sets. Each recording was sampled with a 16 kHz sample rate and transformed into the CQT spectrogram.

\paragraph{Synthesized Instruments}
This work introduces another type of dataset generated from annotations from other datasets. To achieve the best variety of created samples, we took annotations from MAPS and GuitarSet datasets and created purely synthesized datasets using FluidSynth\footnote{http://www.fluidsynth.org/} software and popular soundfont - The Fluid Release 3 General-MIDI Soundfont \footnote{https://member.keymusician.com/Member/FluidR3\_GM/index.html} - used to simulate real instruments. Using this approach, we can generate recordings of many instruments based on existing annotations. For this research, we decided to generate synthesized data from "Acoustic Grand Piano" and "Acoustic Guitar (steel)" MIDI programs of FluidR3\_GM soundfont. Split for training, validation and testing was dataset-wise, which means that data from the training set for MAPS and the training set for GuitarSet were present in the training set for SynthesizedInstruments. We are using only randomly chosen fixed-size sequences of each composition for training, so it should not make this model overfit to traits specific to datasets distributions. Achieved recordings contain the clean version of each instrument, which is often not desired in analyzing noisy real-world data recorded by modern microphones. Each recording generated by the software synthesizer was later sampled with a 16 kHz sample rate and transformed into the CQT spectrogram.

\subsubsection{Data processing}
Each experiment was focused on training on a specific dataset. We used CQT transform as a spectrogram function for all experiments. We used nnAudio library \cite{cheuk2020nnaudio} for spectrogram calculation. To avoid recalculating CQT transform each time, we saved data on disk once it was calculated and loaded it in the subsequent experiments to not perform it again.

\subsubsection{Experimental protocol}
Datasets were split into training, validation, and testing sets. We checked the model using the validation set after every ten learning epochs. After training, all datasets were tested using corresponding testing sets for all available datasets. In the discussion of results, only datasets containing real-world recordings were considered (MAPS and GuitarSet).

\subsubsection{Result analysis}
%TODO - describe metrics - DONE
For model evaluation, we used metrics used by other automatic music transcription works available in mir\_eval library \cite{raffel2014mir_eval} for Python. Standard metrics used for automatic music transcription are precision (P), recall (R), and F1. They are usually measured in different ways described below:
\begin{itemize}
    \item frame - it checks if all notes in the small frame (usually 10ms) are in the correct positions. It may favour models generating interrupted outputs, which sounds undesirable, but it is correct in most frames, and recordings with multiple breaks without sound.
    \item note - a basic metric that checks if the note onset is correct (with small tolerance to start the note sooner or later).
    \item note with offset - it checks not only note onset but also offset, what makes this metric one of the most challenging because many currently existing models tend to recognize the pitch and the onset time correctly, but the note has improper length or output contains interrupted sound.
\end{itemize}

\subsubsection{Implementation and reproducibility}
Experiments were run on a modern PC with Intel(R) Core(TM) i9-10940X CPU @ 3.30GHz, Nvidia GeForce RTX 3090 and 64 GB of RAM. It is essential to mention that depending on the capability of the target hardware user can adjust the sequence length of the audio recording used for training (327680 in our experiments) and the size of the batch (32 in our experiments). Each training was finished after 2000 epochs. Github repository with code for all experiments is available online\footnote{https://github.com/w4k2/automatic\_music\_transcription}

\subsection{Results}
We evaluated each model on different datasets containing real-world recordings. Afterward, we checked the model's behavior during finetuning on weights of Synthesized Instruments and finetuning performed on the model containing weights trained on corresponding datasets.

%\subsubsection{Results for frame metrics}
\begin{table}[h]
    \centering
    \caption{Results for frame metrics}

    \begin{tabular}{|l|c|c|c|c|c|c|}
    \clineB{2-7}{}
    \multicolumn{1}{l}{} & \multicolumn{6}{|c|}{\textsc{Evaluated Datasets}} \\ 
    \clineB{2-7}{}
    \multicolumn{1}{l}{}& \multicolumn{3}{|c|}{\textsc{MAPS}} & \multicolumn{3}{c|}{\textsc{GuitarSet}}     \\
    \hline
    \textbf{Model trained on} & P & R & F1 & P & R & F1     \\
    \hline
    MAPS & 0.745 & \textbf{0.697} & 0.718  & 0.714 & 0.712 & 0.703      \\
    \hline
    GuitarSet & 0.535 & 0.375 & 0.434  & \textbf{0.900} & 0.832 & 0.863      \\
    \hline
    Synthesized Instruments & 0.673 & 0.362 & 0.439  & 0.732 & 0.394 & 0.489      \\
    \hline
    \multicolumn{7}{|c|}{\textsc{Results after transfer from Synthesized Instruments}} \\
    \hline
    Synthesized Instruments -> MAPS & \textbf{0.810} & 0.656 & \textbf{0.722}  & 0.781 & 0.644 & 0.692      \\
    \hline
    SynthesizedInstruments -> GuitarSet & 0.533 & 0.412 & 0.460  & 0.876 & 0.873 & \textbf{0.873}      \\
    \hline
    \multicolumn{7}{|c|}{\textsc{Results after transfer from GuitarSet}} \\
    \hline
    GuitarSet -> MAPS & 0.765 & 0.656 & 0.704  & 0.710 & 0.698 & 0.692      \\
    \hline
    \multicolumn{7}{|c|}{\textsc{Results after transfer from MAPS}} \\
    \hline
    MAPS -> GuitarSet & 0.528 & 0.424 & 0.466  & 0.860 & \textbf{0.884} & 0.870      \\
    \hline
    
    \end{tabular}
    \label{tab:frame_results}
\end{table}
In metrics checking the entire frame of output piano roll, the best result for precision and F1 for MAPS dataset was achieved similarly for the model finetuned on MAPS dataset after transfer of weights from the model trained on Synthesized Instruments. The model trained only on MAPS samples has a greater recall value, suggesting that it generates too much output. For the GuitarSet dataset, it appears that using Synthesized Instruments as a base did not improve precision and recall for transcription, but F1 is greater than for any other training. Detailed results for \textit{frame} metrics are presented in table \ref{tab:frame_results}.  

%\subsubsection{Results for note onset metrics}
\begin{table}[h]
    \centering
    \caption{Results for note metrics}
    \begin{tabular}{|l|c|c|c|c|c|c|}
    \clineB{2-7}{}
    \multicolumn{1}{l}{} & \multicolumn{6}{|c|}{\textsc{Evaluated Datasets}} \\ 
    \clineB{2-7}{}
    \multicolumn{1}{l}{} & \multicolumn{3}{|c|}{\textsc{MAPS}} & \multicolumn{3}{c|}{\textsc{GuitarSet}}     \\
    \hline
     \textbf{Model trained on}:  & P & R & F1 & P & R & F1     \\
    \hline
    MAPS & 0.630 & 0.663 & 0.642  & 0.615 & 0.708 & 0.646      \\
    \hline
    GuitarSet & 0.232 & 0.234 & 0.229  & \textbf{0.795} & 0.733 & \textbf{0.754}      \\
    \hline
    Synthesized Instruments & 0.494 & 0.414 & 0.423  & 0.589 & 0.534 & 0.537      \\
    \hline
    \multicolumn{7}{|c|}{\textsc{Results after transfer from Synthesized Instruments}} \\
    \hline
    SynthesizedInstruments -> MAPS & \textbf{0.655} & \textbf{0.690} & \textbf{0.669}  & 0.678 & \textbf{0.754} & 0.701      \\

    \hline
    SynthesizedInstruments -> GuitarSet  & 0.215 & 0.250 & 0.227  & 0.751 & 0.747 & 0.740      \\
        \hline
    \multicolumn{7}{|c|}{\textsc{Results after transfer from GuitarSet}} \\
    \hline
    GuitarSet -> MAPS & 0.624 & 0.682 & 0.648  & 0.588 & 0.741 & 0.642      \\
        \hline
    \multicolumn{7}{|c|}{\textsc{Results after transfer from MAPS}} \\
    \hline
    MAPS -> GuitarSet & 0.242 & 0.239 & 0.236  & 0.767 & 0.733 & 0.740      \\
    \hline
    \end{tabular}
    \label{tab:note_results}
\end{table}
In metrics checking the onset of note, the best result for MAPS dataset was achieved for the model finetuned on MAPS dataset after the weight transfer from the model trained on Synthesized Instruments. For GuitarSet dataset, it appears, that using Synthesized Instruments as a base did not improve precision and F1 for transcription, but recall is greater than for any other training. It may suggest that the model creates more outputs based on spectrograms, but it does not improve transcription results. Detailed results for \textit{note} metrics are presented in table \ref{tab:note_results}.  

%\subsection {Results for note with offset metrics}
\begin{table}[]
    \centering
    \caption{note-with-offset results}

    \begin{tabular}{|l|c|c|c|c|c|c|}
    \clineB{2-7}{}
    \multicolumn{1}{l}{} & \multicolumn{6}{|c|}{\textsc{Evaluated Datasets}} \\ 
    \clineB{2-7}{}
    \multicolumn{1}{l}{}& \multicolumn{3}{|c|}{\textsc{MAPS}} & \multicolumn{3}{c|}{\textsc{GuitarSet}}     \\
    \hline
    \textbf{Model trained on} & P & R & F1 & P & R & F1     \\
    \hline
    MAPS & 0.406 & 0.428 & 0.414  & 0.246 & 0.298 & 0.265      \\
    \hline
    GuitarSet & 0.062 & 0.061 & 0.060  & \textbf{0.583} & 0.554 & 0.564      \\
    \hline
    SynthesizedInstruments & 0.236 & 0.191 & 0.198  & 0.102 & 0.104 & 0.100      \\
    \hline
    \multicolumn{7}{|c|}{\textsc{Results after transfer from Synthesized Instruments}} \\
    \hline
    Synthesized Instruments -> MAPS & \textbf{0.423} & \textbf{0.445} & \textbf{0.432}  & 0.221 & 0.260 & 0.236      \\
    \hline
    SynthesizedInstruments -> GuitarSet & 0.063 & 0.072 & 0.066  & 0.547 & 0.556 & 0.548      \\
    \hline
    \multicolumn{7}{|c|}{\textsc{Results after transfer from GuitarSet}} \\
    \hline
    GuitarSet -> MAPS & 0.392 & 0.428 & 0.407  & 0.236 & 0.312 & 0.264      \\
    \hline
    MAPS -> GuitarSet & 0.071 & 0.070 & 0.069  & 0.574 & \textbf{0.566} & \textbf{0.566}      \\
    \hline
    \end{tabular}
    \label{tab:note_with_offset_results}
\end{table}
In metrics checking notes with their corresponding offsets, the best result for MAPS dataset was achieved by the model finetuned on MAPS dataset after the weight transfer from the model trained on Synthesized Instruments. For the GuitarSet dataset, using MAPS model as a base improved recall and F1 for guitar transcription, but precision is a little bit better for the model trained on GuitarSet. Detailed results for \textit{note-with-offset} metrics are presented in table \ref{tab:note_with_offset_results}.  

\subsection {Time to achieve model convergence}
\begin{figure}[ht]
    \centering
    \includegraphics[width=13cm]{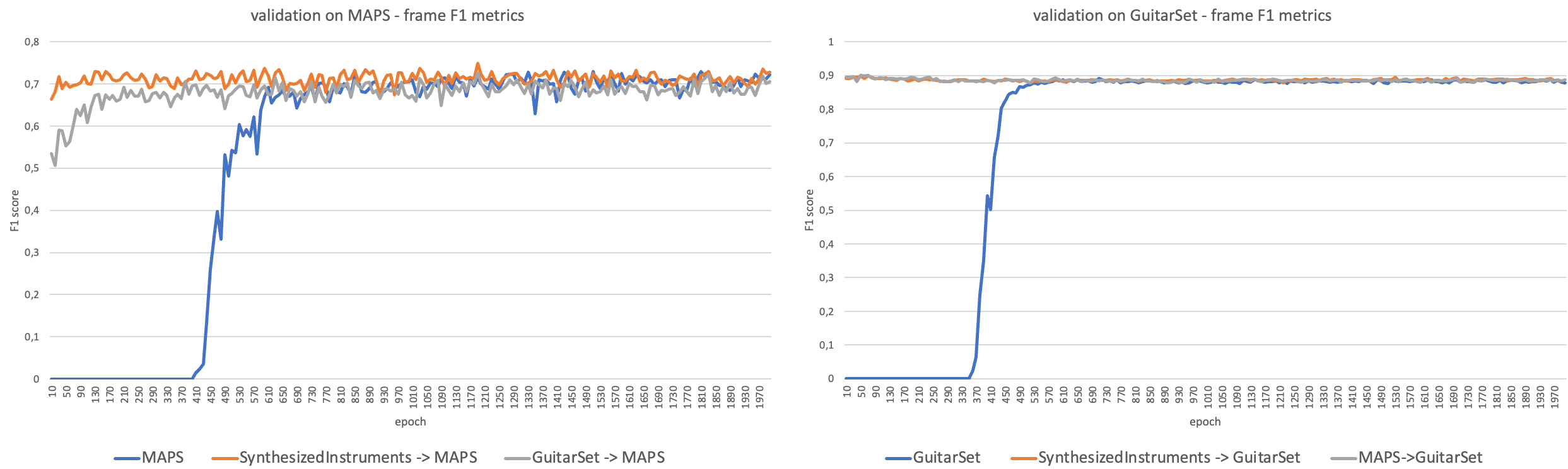}
    \caption{Validation on target datasets during training of models.}
    \label{fig:validation_during_training}
\end{figure}
%TODO add figure with comparison of speed of adaptation to next instrument - DONE
During experiments we noticed, that for finetuning on model trained on SynthesizedInstruments metrics achieved satisfying results significantly faster than on finetuning of piano on weights trained for guitar-only recordings or learning model from random weights. This interesting capability may be used to quickly adapt existing automatic music transcription models for another type of instruments. It is visualized on figure \ref{fig:validation_during_training}. It is especially visible for finetuning to MAPS dataset. Learning from randomly initialized weights needed about 600 epochs to achieve good results in measured metrics. When we tried to finetune model trained on Guitarset it needed a similar time to achieve the same results for MAPS metrics as the model trained on Synthesized Instruments. For GuitarSet data we can see, that both pretraining on MAPS and Synthesized Instruments gave similar results from the first epoch. It indicates that model trained on synthesized data is very good candidate to perform transfer and quickly achieve well results for different timbres.

\subsection{Lessons learned}
%TODO write more, answer research questions - DONE
% \begin{itemize}
%     \item RQ1: How does the U-net model trained on synthesized data perform on real-world recordings?
%     \item RQ2: How does knowledge transfer from the U-net model trained on synthesized data impact training on real-world recordings?
%     \item RQ3: Is the U-net model trained on synthesized data a better candidate for fine-tuning on other instruments than the U-net model trained on real-world piano recordings?
% \end{itemize}
We may claim from the results that the model trained on Synthesized Instruments performed well for MAPS and GuitarSet even after zero-shot transfer. It can perform transcription on real-world recordings without seeing any real-world sample (RQ1 answered). After transfer from Synthesized Instruments, we showed that both GuitarSet and MAPS datasets are easily adjustable to the target domain. The time to achieve satisfying results in MAPS evaluation is significantly better for Synthesized Instruments than for GuitarSet, which gives hope for such a model's usefulness to create adjustable submodels quickly focused on specific instruments (RQ2 answered). Finetuning after initialization with weights trained for Synthesized Instruments on the MAPS achieved the best result for this dataset. Finetunning on GuitarSet real-world recordings of the model learned on MAPS seems to perform better in most cases. It may be related to the fact that the model trained on the MAPS dataset learned to handle microphone-originated noises better. From the results of experiments, a model well-trained on real-world piano recordings may be a better candidate to transfer knowledge to a model focused on other instruments (RQ3 answered). However, GuitarSet is a small dataset, and more experiments with different real-world instrument recordings and samples may be needed to assess the usefulness of the training in the Synthesized Instruments domain.

\section{Conclusion}
%TODO [WAIT till the end]
We presented that the synthesized audio data may be a valuable possible source of knowledge to train models for real-world music data. Artificially created different timbres of instruments allow the model to generalize other instrument outputs and perform well on real-world data after zero-shot training. This model cannot directly compete with a model trained on diverse real-world data containing many features related to environmental conditions like different sounds in recording and microphone-specific noises. Transfer of knowledge at the beginning of training allows us to quickly create models with satisfying metrics values, which may be especially useful for problems where the different distributions occur over time. After a long training session, applying knowledge from the model trained on synthesized data may improve existing metrics for more complicated datasets. It may be a good reason for making prototypes for another kind of models focused on resolving automatic music transcription.

Further research may focus on the impact of adding further synthesized instruments to experiment on automatic music transcription generalization and to check the influence of input modification realized by special audio filtering on output. It may be worth reviewing domain-specific spectrogram changes related to different timbres, which could extract special traits specific to each timbre. Finding accurate solutions for spectrogram representation for different instruments is essential to ensure that further research will overcome problems with music transcription generalization.

\section*{Acknowledgment}
This work is supported by the CEUS-UNISONO programme, which has received funding from the National Science Centre, Poland under grant agreement No. 2020/02/Y/ST6/00037.

We would like to thank Jędrzej Kozal for his support during creation of this work.
%To be added if paper is accepted.
\bibliographystyle{splncs04}
\bibliography{sn-bibliography}

\end{document}